\documentclass[showpacs,prb,preprint]{revtex4}
\usepackage{amsmath}
\usepackage{amssymb}
\usepackage{graphicx}

\begin{document}

\title{Weak ferromagnetism of antiferromagnetic domains in graphene with
defects}
\author{Y. G. Semenov, J. M. Zavada, and K. W. Kim}

\begin{abstract}
Magnetic properties of graphene with randomly distributed magnetic
defects/vacancies are studied in terms of the Kondo Hamiltonian in the mean
field approximation. It has been shown that graphene with defects undergoes
a magnetic phase transition from a paramagnetic to a antiferromagnetic (AFM)
phase once the temperature reaches the critical point $T_{N}$. The defect
straggling is taken into account as an assignable cause of multiple
nucleation into AFM domains. Since each domain is characterized by partial
compensating magnetization of the defects associated with different
sublattices, together they reveal a super-paramagnetic behavior in a
magnetic field. Theory qualitatively describe the experimental data provided
the temperature dependence of the AFM domain structure.
\end{abstract}


\address{Department of Electrical and Computer Engineering, North Carolina
State University, Raleigh, NC 27695-7911}




\maketitle
\bigskip

Along with unique transport characteristics the magnetic behavior of the
graphene-based materials attracts much attention in recent studies due to
the significant interest in fundamental physics and prospective spintronic
applications. Particularly, the possibility of the band gap manipulating in
antiferromagnetic ordered defective graphene (i.e. graphene with adatoms or
vacancies)\cite{Daghofer10,Rappoport10,Balog10} offers an additional control
for nonlinear functionality while an efficient spin injection capability
into graphene\cite{Kawakami10} with long spin coherence time/length even at
room temperature \cite{Tombros07} puts the graphene in the forefront of the
materials for emerging spin-based information processing. Furthermore, the
room temperature weak ferromagnetism (FM) has been reported in highly
oriented pyrolitic graphite irradiated by proton beams\cite{Esquinazi03} and
in defective graphene prepared from soluble functionalized graphene sheets.%
\cite{Wang09} Moreover, it was experimentally discovered the coexist of
ferromagnetic correlations along with antiferromagnetic (AFM) interactions
in all series of the multilayer defective graphene samples in Ref.~%
\onlinecite{Matte09}. On the other hand, no ferromagnetism has been detected
in pure graphene nanocrystals in wide range of temperatures~\cite{Sepioni10}
revealing ambiguity about the graphene edge contribution to magnetization
data (compare Refs.~\onlinecite{Yazyev08,Cervenka09,Lin09}).

It was recently theoretically and experimentally realized that the vacancies
or hydrogen addatoms associated with carbon atoms mediate the local magnetic
moments in the graphene.\cite%
{Yazyev07,Kumazaki07,Pisani08,Palacios08,Yazyev10} Moreover, the local spin
moments of the defects reveal strong exchange interaction with delocalized
electrons that is the main source to trigger low temperature anomaly in
conductivity and Kondo effect.\cite{Hentschel07,Zhu10,Fuhrer10,Uchoa10}
Theoretical analysis of indirect interaction through the graphene carriers
in terms of Kondo Hamiltonian unambiguously indicates the ferromagnetic
(antiferromagnetic) exchange coupling between localized spins associated
with the same (different) sublattices of the crystalline graphene.\cite%
{Brey07,Cheianov07,Rappoport09,Venezuela09,Schaffer10} This fact as well as
Lieb's theorem\cite{Leib} are usually adduced to back up the arguments that
a weak ferromagnetism is a result of imbalance $\Delta N_{d}$ in the number
of vacancies/impurities of the A or B sublattices (i.e. not precise
compensation of sublattices magnetizations with opposite directions).\cite%
{Easquinazi10,Lopez10} At the same time the actual imbalance of randomly
distributed spin moments fails even a weak ferromagnetism because $%
\left\vert \Delta N_{d}\right\vert /N_{d}\rightarrow 0$ in large graphene
crystal with the total number $N_{d}$ of the defects.

In present study we propose a different approach to the problem of
coexistence of FM and AFM phase in a single graphene layer. It takes into
account a multiple nucleation of antiferromagnetic domain germs in the mass
of randomly distributed spins in monolayer graphene. At the same time, the
domains each reveal a small random imbalance $\Delta n_{d}$ and magnetism so
that together they \textit{additively} contribute to the net magnetic moment
culminating the saturated magnetization in a magnetic field. In such a way
the finiteness of the magnetic correlations should be taken into account.
The strong short-range correlation within each domain is responsible for its
antiferromagnetic ordering. Such AFM correlations are weakened or broken at
the domain boundaries, which can be associated with the impurity
rarefactions or lattice imperfects. Thus the behavior of the magnetic
domains in a magnetic field is almost mutually independent that constitutes
some finite magnetization in the limit $N_{d}\rightarrow \infty $. It is
important to note that magnetism in graphene does not come into conflict
with Mermin-Wagner theorem \cite{Mermin66} because of finite magnetic
anisotropy \cite{Brataas06,Yazyev08} and restricted sizes of AFM domains.

The microscopic approach is based on the Kondo Hamiltonian of
carrier-localized spin exchange interaction that has been extensively used
in the analysis of magnetic properties of graphene in Refs.~%
\onlinecite{Hentschel07,Uchoa10,Rappoport10} and \onlinecite{Schaffer10}.
The spectrum of bulk graphene is treated in tight binding approximation. In
mean field approximation, this approach leads to set of two equations for
the sublattice magnetizations that define a critical temperature of the
antiferromagnetic ordering without any limitations on the number of carbon
atoms involving to computation in more sophisticated models (Refs.~%
\onlinecite{Daghofer10} and \onlinecite{Rappoport10}). As a result, we find
a simple analytical expression for N\'{e}el temperature that may serve as a
guide for analysis of the numerous experimental data on graphene magnetism
and estimate the mean size of the magnetic domains in graphene.

Let us consider a graphene fragment possessed large enough a flat area $%
A_{f} $ to neglect the edge effects. The Hamiltonian in momentum
representation takes the form \cite{Neto09}%
\begin{equation}
H_{\mathbf{k}}=\gamma _{cc}[\widehat{\sigma }_{1}f_{1}(\mathbf{k})+\widehat{%
\sigma }_{2}f_{2}(\mathbf{k})],  \label{1}
\end{equation}%
where $\gamma _{cc}=2.7$ eV is the matrix element of electron hopping
between nearest neighbor atoms connected with the vectors $\mathbf{e}_{m}$ ($%
m$=1,2,3), the Pauli matrixes $\widehat{\sigma }_{i}$ are defined over the
sublattices $A$ and $B$ basic functions and $\mathbf{k=}(k_{x},k_{y})$ is
the electron momentum; $f_{1}(\mathbf{k})=\sum_{m}\cos (\mathbf{k\mathbf{e}}%
_{m})$, $f_{2}(\mathbf{k})=\sum_{m}\sin (\mathbf{k\mathbf{e}}_{m})$. We
assign $\mathbf{e}_{m}=a_{cc}(\cos m\omega ,\sin m\omega )$, where $%
a_{cc}=0.142$ nm and $\omega =2\pi /3$. In diagonal form, Hamiltonian (\ref%
{1}) describes the graphene dispersion law%
\begin{equation}
\epsilon _{b,\mathbf{k}}=\gamma _{cc}b\varepsilon (a_{0}\mathbf{k})
\label{9}
\end{equation}%
for conduction ($b=1$) and valence ($b=-1$) bands with $\varepsilon (a_{0}%
\mathbf{k})=\sqrt{f_{1}^{2}(\mathbf{k})+f_{2}^{2}(\mathbf{k})}$. A
straightforward algebra shows that $\varepsilon (\mathbf{q})=\left( 3+4\cos
\frac{\sqrt{3}q_{x}}{2}\cos \frac{q_{y}}{2}+2\cos q_{y}\right) ^{1/2}$, i.e.
there are two non-equivalent contact points at the corners of first
Brillouin zone (BZ) $\mathbf{K,K}^{\prime }=\frac{\omega }{a_{0}}(\sqrt{3}%
,\pm 1,)$ where conduction and valence bands are degenerated at $\epsilon
_{b,\mathbf{K}}=\epsilon _{b,\mathbf{K}^{\prime }}=0$; $a_{0}=\sqrt{3}a_{cc}$
is a length of lattice vectors. In the vicinity of these points the energy
dispersions are the Dirac cones, $\epsilon _{b,\mathbf{K}^{(\prime )}+%
\mathbf{\varkappa }}=b\hbar v_{F}\left\vert \mathbf{\varkappa }\right\vert $%
, with Fermi velocity $v_{F}=\frac{\sqrt{3}}{2}\gamma _{cc}a_{0}/\hbar $.

The Kondo Hamiltonian of the exchange interaction between a band electron
(with position $\mathbf{r}$ and spin $\mathbf{S}$) and the $n_{d}$ localized
spin moments $\mathbf{I}_{j}$ pinned to the sites $\mathbf{R}_{j}$ of the
graphene lattice reads
\begin{equation}
H_{K}=-\sum_{j=1}^{n_{d}}J(\mathbf{r},\mathbf{R}_{j})\mathbf{I}_{j}\mathbf{S,%
}  \label{v2}
\end{equation}%
where $J(\mathbf{r},\mathbf{R}_{j})\approx J\varpi _{0}\delta (\mathbf{r}-%
\mathbf{R}_{j})$, $J$ is the exchange constant, $n_{d}=n_{A}+n_{B}$ the
total number of the defects located at $A$ and $B$ sublattices, $\varpi _{0}=%
\sqrt{3}a_{0}^{2}/2$. In the bipartite graphene lattice, it is convenient to
double group the Hamiltonian of Eq.~(\ref{v2}) on sublattice defects $%
j_{A(B)}=1,...,$ $n_{A(B)}$: $H_{K}=H_{K}^{(A)}+H_{K}^{(B)}$. In the
representation of eigenfunctions of Eq.~(\ref{1}), each part of $H_{K}$
manifests itself through the projection operators $P_{A(B)}=(1\pm \widehat{%
\sigma }_{3})/2$. Then the summing up over the random scattered impurities
and thermal averaging of their spin states reduce Eq.~(\ref{v2}) to%
\begin{equation}
H_{ex}=\alpha \mathbf{(m\mathbf{S+w}C),}  \label{v5a}
\end{equation}%
where the total magnetic moment of the defects $\mathbf{m}=\mathbf{m}_{A}+%
\mathbf{m}_{B}$ and their antiferromagnetic vector $\mathbf{w}=\mathbf{m}%
_{A}-\mathbf{m}_{B}$ are expressed via sublattice magnetic moments $\mathbf{m%
}_{A(B)}=-n_{A}g\mu _{B}\left\langle \mathbf{I}_{j_{A(B)}}\right\rangle $; $%
\alpha =\frac{1}{n_{f}}\frac{J}{g\mu _{B}}$, $n_{f}=A_{f}/\varpi _{0}$ the
total number of primitive sells, $\varpi _{0}$ is their area, $g\cong 2$ and
$\mu _{B}$ are the $g$-factor and Bohr magneton. The distinguishing property
of Eq.~(\ref{v5a}) is that AFM vector exerts the composite spin $\mathbf{C}=%
\widehat{\sigma }_{3}\otimes \mathbf{S}$ that can be a finite magnitude even
under $\left\langle \mathbf{S}\right\rangle =0$.

Let us introduce the effective magnetic fields $\mathbf{B}_{A}$ and $\mathbf{%
B}_{B}$ that cause the spin polarizations in each sublattice $\left\langle
\mathbf{I}_{j_{A(B)}}\right\rangle =-\mathbf{n}_{A(B)}\frac{\mathbf{1}}{2}%
\tanh \frac{g\mu _{B}B_{A(B)}}{2T}$, $\mathbf{n}_{A(B)}=\mathbf{B}%
_{A(B)}/B_{A(B)}$, $k_{B}=1$. On the other hand they can be represented as
\cite{LP9}
\begin{equation}
\mathbf{B}_{A(B)}\mathbf{=}\mathbf{-}\frac{\partial \Phi }{\partial \mathbf{m%
}_{A(B)}}  \label{e3}
\end{equation}%
in terms of of thermodynamic potential
\begin{equation}
\Phi =-T\sum_{b,s,\mathbf{k}}\ln \left( 1+e^{(\mu -E_{b,s,k})/T}\right) ,
\label{e2}
\end{equation}%
and chemical potential $\mu $. The energy bands of the defective graphene
with Hamiltonian $H_{\mathbf{k}}+H_{ex}$ are described by
\begin{equation}
E_{b,s,\mathbf{k}}=\frac{s}{2}\alpha m+b\sqrt{\epsilon _{b,\mathbf{k}%
}^{2}+\left( \frac{\alpha w}{2}\right) ^{2}},  \label{e8}
\end{equation}%
where $s=\pm 1$ is a spin number. A distinctive feature of the dispersion
law (\ref{e8}) consists in opening bandgap $E_{g}=\alpha (w-m)$ provided
that $w>m$. Such splitting of the energy bands lowers the total electronic
energy of the valence band that cannot be compensated by raising electron
energy in the conduction band resembling the cooperative Jahn-Teller effect.%
\cite{Englman72} Equation~(\ref{e3}) constitutes the closed set of the
equations for both $\mathbf{B}_{A}=\widehat{\mathbf{z}}B_{A}$ and $\mathbf{B}%
_{B}=\widehat{\mathbf{z}}B_{B}$, the unit vector $\widehat{\mathbf{z}}$ is
directed along quantization axis. Explicitly they take the form of integral
over the first BZ,
\begin{eqnarray}
B_{A,B} &\mathbf{=}&\mathbf{-}\frac{\varpi _{0}}{g\mu _{B}(2\pi )^{2}}\int
\int_{BZ}dq_{x}dq_{y}F(\mathbf{k});  \label{e5} \\
F(\mathbf{k}) &=&J\sum_{b,s}\left( \frac{s}{2}\pm \frac{b\alpha w}{4\sqrt{%
\epsilon _{b,\mathbf{k}}^{2}+\left( \frac{\alpha w}{2}\right) ^{2}}}\right)
f(E_{b,s,\mathbf{k}}-\mu ),  \label{e6}
\end{eqnarray}%
where $f(E_{b,s,\mathbf{k}}-\mu )$ is the Fermi-Dirac function and $\pm $
discriminates the different sublattices.

In the first stage we focus on the stronger effect of AFM ordering within a
single domain that eventually establishes weak FM. Ignoring small imbalance $%
\Delta n_{d}\ll n_{d}$, Eqs.~(\ref{e5}) and (\ref{e6}) decompose on
independent equations for $B_{A}$ and $B_{B}=-B_{A}$ so that each $B_{A}$
and $B_{B}$ satisfy to the self-consistent equation
\begin{equation}
y=\frac{J^{2}x}{T\gamma _{cc}}I(y,\mu ,T)\tanh \left( \frac{y}{2}\right) ,
\label{y1}
\end{equation}%
in terms of variable $y=g\mu _{B}B_{A(B)}/T$. A complex integral function $%
I(y,\mu ,T)$ can be reduced with high accuracy to the constant $I_{0}\simeq
0.448$ provided that $\gamma _{cc}\gg Jx,\mu ,T$. Therewith in the limit $%
y\rightarrow 0$ Eq.~(\ref{y1}) gives rise the expression for N\'{e}el
temperature
\begin{equation}
T_{N}=\frac{J^{2}x}{2\gamma _{cc}}I_{0},  \label{tn}
\end{equation}%
which depends merely on the exchange constant and defect fraction $%
x=n_{d}/2n_{f}$. The squared dependence of $T_{N}$ on $J$ as well as the
quantitative evaluation $T_{N}=0.0112\gamma _{cc}$ at $x=0.2$ and $%
J=0.5\gamma _{cc}$ are in a good agreement with Monte-Carlo simulations
carried out in Ref. \onlinecite{Rappoport10}. Particulary, the Eq.(\ref{tn})
shows that exchange constant $\left\vert J\right\vert =1.9$ eV guarantees a
room temperature AFM ordering at reasonable low $x=0.04$. Solutions $%
y=y(x,T) $ of Eq.~(\ref{y1}) define also the energy gap $E_{g}=Jx\tanh \frac{%
y(x,T)}{2}$ and magnetization $m_{A}=\frac{1}{2}n_{A}g\mu _{B}\tanh \frac{%
y(x,T)}{2}$. Fig. 1 illustrate the temperature dependence of $E_{g}$
calculated with Eq.~(\ref{y1}).

Nucleation of $N_{D}$ AFM domains each in possess of a random imbalance $%
n_{A}-n_{B}$ lays the groundwork for weak ferromagnetism in graphene.
Applying the statistical approach, let us attribute $x$ to the equal
probability for each site $A$ or $B$ to be defected for all $N_{D}$ domains
so that mean number of deviation is $\overline{(n_{A}-n_{B})}=0$ because $%
\overline{n_{A}}=\overline{n_{B}}=xn_{f}$. The mean-square estimate is $%
\overline{(n_{A}-n_{B})^{2}}=2n_{f}x(1-x)=n_{d}(1-x)$, thus the mean number
of non-compensated spins is$\overline{\text{ }\Delta n_{d}}=\sqrt{%
2n_{f}x(1-x)}=\sqrt{n_{d}(1-x)}$. The respective magnetic moment
\begin{equation}
m(n_{f},x,T)\simeq \mu _{B}\sqrt{2n_{f}x(1-x)}\tanh \frac{y(x,T)}{2}
\label{md}
\end{equation}%
can be interpreted as a weak ferromagnetism attributed to a single domain.

Above we considered the formation of the magnetic moments $m(n_{f},x,T)\leq $
$\mu _{B}\sqrt{n_{d}}$ ($n_{d}\gg \sqrt{n_{d}}\gg 1$) in individual AFM
domains. To consider the behavior of multi-domain graphene in a magnetic
field $\mathbf{B}_{0}$ further conjectures must be done. It is convenient to
split the ensemble of domains on subsets, each determined by certain space $%
A_{f}$ and number $n_{d}$ of the defects. In a magnetic field $\mathbf{B}%
_{0} $ the $A_{f}$ -domain with $n_{d}$ defects contributes to the net
magnetization as $m(n_{f},x,T)L[B_{0}m(n_{f},x,T)/T_{eff}]$, where $L(x)$ is
a Langevin function and $T_{eff}=T+T_{AF}$ is an effective temperature.
Parameter $T_{AF}$ takes into account the tendency to merge several small
domains into one large AFM one that reveals a close analogy of this
parameter with the temperature shift $T_{0}$ in diluted magnetic
semiconductors with AFM interaction between localized spin moments.\cite%
{Bednarski00} Such interdomain interaction constitutes proportionality of $%
T_{AF}$ to the length of domain boundary network or inverse proportionality
to domain mean size $L$. The final magnetization output $M$ can be expressed
in terms of distribution function $f(A_{f},x)$ so that
\begin{equation}
M=\int \int m(n_{f},x,T)L\left( \frac{B_{0}m(n_{f},x,T)}{T_{eff}}\right)
f(A_{f},x)dA_{f}dx.  \label{mm}
\end{equation}

Apparently the dispersion in the domain shapes and spaces and in the defect
densities is very specific for particular sample preparation, chemical and
thermal treatment; therefore the $f(A_{f},x)$ cannot be specified \textit{a
priori}. So in the rest part of the paper we focus on the analysis of
particular experimental data on graphene magnetization reported in Ref. %
\onlinecite{Wang09}. There was found that at room temperature, a magnetic
field $B_{0}\approx 3$ kOe saturates magnetization at small amounts $%
M_{s}=0.02$ emu/g and $M_{s}=0.004$ emu/g in two different graphene samples
Gr600 and Gr400 respectively. At the same time a stronger magnetic field $%
B_{0}\approx 30$ kOe needs to saturate magnetization with much higher $%
M_{s}=0.8$ emu/g and $M_{s}=0.2$ emu/g at $T=2$ K. Besides, in very narrow
magnetic field region there was recorded some hysteresis loop stipulated by
a weak spin-orbital coupling. This effect is unobtrusive in the scale of
magnetization curves recorded in Ref. \onlinecite{Wang09} and is not
discussed here.

Two different scenarios can be applied to these results. First, let us
assume that the $f(A_{f},x)$ is unchangeable in all range of the
temperatures. In such a case the experiment implies that the majority of the
domains possess relatively small area with slight magnetic moments $%
m(n_{f},x,T)$ and relatively low susceptibility at $T=2$ K. With temperature
increase this portion of the domains undergos to phase transition to
paramagnetic state and drop out from the consideration. The rest part of
large domains with higher magnetic susceptibility can be responsible for
magnetization at $T=300$ K. This scenario, however, fails to describe the
temperature variations of the curves $M=M(B_{0})$ predicting $M_{s}$ to be
much smaller than the observable magnitudes at $T=300$ K.

Another approach assumes that the parameters $n_{f}=A_{f}/\varpi _{0}$ and $%
n_{d}=2xn_{f}$ obey to normal (Gaussian) distributions $G(n_{f})$ and $%
G(n_{d})$ around their mean values $\overline{n_{f}}$ and $\overline{n_{d}}$
but $\overline{n_{f}}$ can vary with $T$. Substituting $%
f(A_{f},x)dA_{f}dx=f(n_{f},n_{d})dn_{f}dn_{d}$, where $%
f(n_{f},n_{d})=N_{D}G(n_{d})G(n_{f})$ into Eq.~(\ref{mm}) allows to
describe all experimental data with good accuracy (Fig. 2). At the
same time one must assume a growth of the domain mean sizes $L=$
$\overline{A_{f}}^{1/2}$ with temperature. Note that the related
effect of $T_{AF}$ decrease as $L^{-1}$(as was discussed above)
correlates with this model (Fig. 2). Such variations as well as
detail domain structure are caused by structure inhomogeneous that
is beyond the developed theory. At the same time, the strengthening
of AFM correlations through the domain boundaries with temperature
increase seems not surprising if high temperatures favors to
electron overcome the barriers between different domains. Note also
that the intriguing result of vanishing ferromagnetism in the sample
\cite{Wang09} Gr800 might just be the effect of inhomogeneous
removing after high temperature annealing that decreases magnetization as $%
A_{f}^{-1/2}\rightarrow 0$.

In conclusion, we have shown that the imbalance in the numbers of the
defects located at $A$ and $B$ sublattices along with graphene fragmentation
into the AFM domains results in small but finite magnetization response in a
magnetic field. Theory quantitatively describe experimental data provided
that the domain structure can vary with temperature. The following
experimental verification needs to establish the model applicability for
particular samples of the defective graphene.

This work was supported in part by the US Army Research Office and the FCRP
Center on Functional Engineered Nano Architectonics (FENA). JMZ acknowledges
support from NSF under the IR/D program.

\newpage

\begin{center}
\begin{figure}[tbp]
\includegraphics[scale=0.7,angle=0]{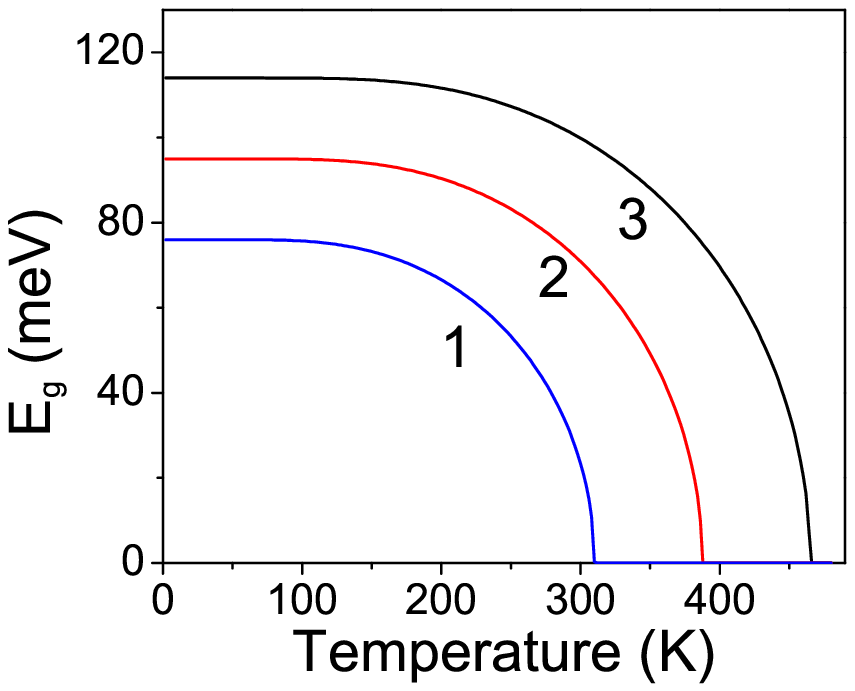}
\caption{Band gap of the defective graphene vs temperature at $J=1.9$ eV and
defect molar fractions 4\%, 5\% and 6\% (curves 1, 2 and 3 respectively).}
\end{figure}
\end{center}

\newpage

\begin{center}
\begin{figure}[tbp]
\includegraphics[scale=0.7,angle=0]{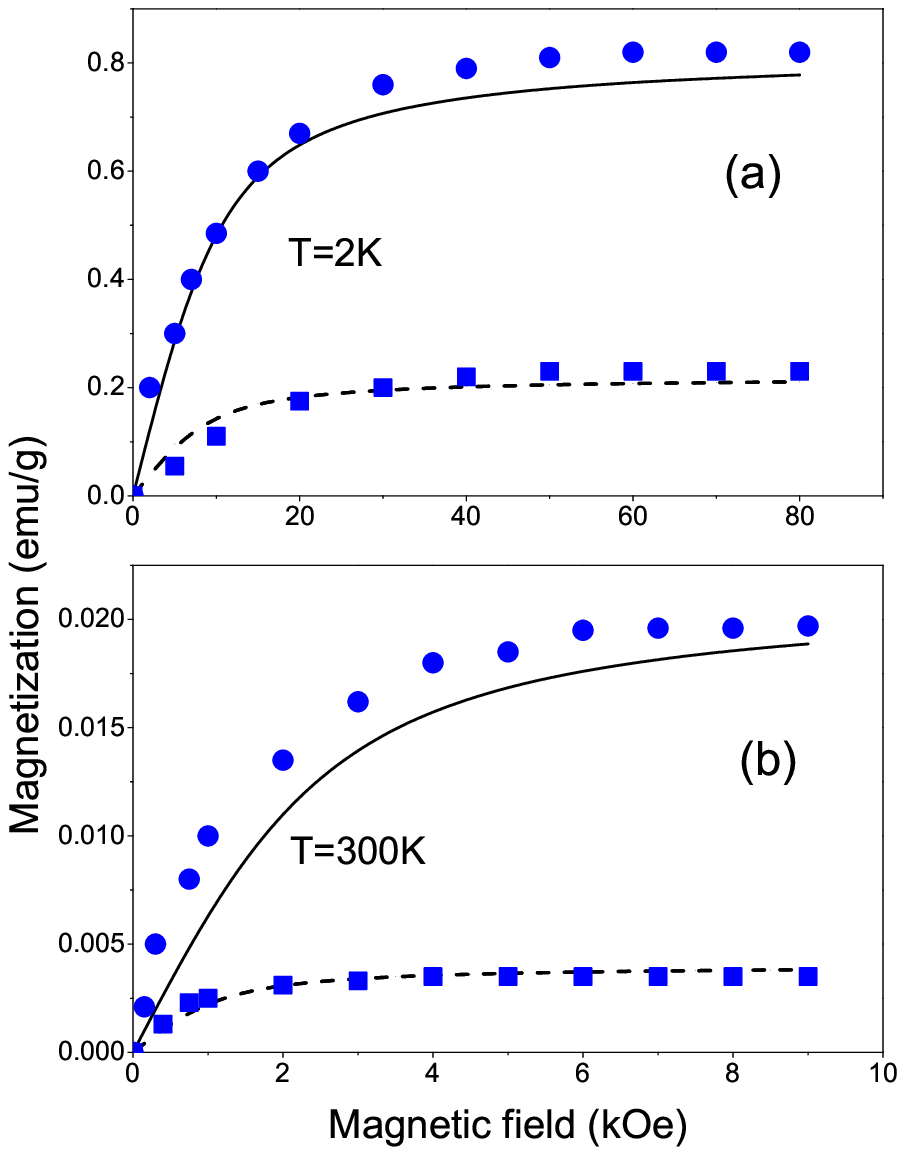}
\caption{Magnetization curves of the defective graphene calculated with Eq. (%
\protect\ref{mm}) (solid and dashed lines). Experimental data are taken from
Ref. \onlinecite{Wang09} for the samples Gr600 (circles) and Gr400 (squares)
for T=2 K (a) and T=300 K (b). Fitting to the Gr600 data provides the $%
x=0.19 $ and mean sizes of the domains $L=43$ nm ($T=2$ K, $T_{AF}=58$ K)
and $L=1.6 $ $\protect\mu$m ($T=300$ K, $T_{AF}=1.5$ K); respectively $%
x=0.12 $, $L=130$ nm ($T=2$ K, $T_{AF}=18$ K) and $L=5.6$ $\protect\mu$m ($%
T=300$ K, $T_{AF}=0.5$ K) have been applied to Gr400.}
\end{figure}
\end{center}


\clearpage

\begin{thebibliography}{99}
\bibitem{Daghofer10} M. Daghofer, N. Zheng, and A. Moreo, Phys. Rev. B
\textbf{82}, 121405(R) (2010).

\bibitem{Rappoport10} T. G. Rappoport, M. Godoy, B. Uchoa, R. R dos Santos,
and A. H. Castro Neto, arXive:1008.3189v1.

\bibitem{Balog10} R. Balog, B. J\o rgensen, L. Nilsson, M. Andersen, E.
Rienks, M. Bianchi, M. Fanetti, E. L\ae gsgaard, A. Baraldi, S. Lizzit, Z.
Sljivancanin, F. Besenbacher, B. Hammer, T. G. Pedersen, P. Hofmann, and L.
Hornek{\ae }r, Nature Mat. \textbf{9}, 315 (2010).

\bibitem{Kawakami10} W. Han, K. Pi, K. M. McCreary, Y. Li, J. J. I. Wong, A.
G. Swartz, and R. K. Kawakami, Phys. Rev. Lett. \textbf{105}, 167202 (2010).

\bibitem{Tombros07} N. Tombros, C. Jozsa, M. Popinciuc, H. T. Jonkman, and
B. J. van Wees, Nature \textbf{448}, 571 (2007).

\bibitem{Esquinazi03} P. Esquinazi, D. Spemann, R. H\"{o}hne, A. Setzer,
K.-H. Han, and T. Butz, Phys. Rev. Lett. \textbf{91}, 227201 (2003).

\bibitem{Wang09} Y. Wang, Y. Huang, Y. Song, X. Zhang, Y. Ma, J. Liang, and
Y. Chen, Nano Lett. \textbf{9}, 220 (2009).

\bibitem{Matte09} H. S. S. R. Matte, K. S. Subrahmanyam, and C. N. R. Rao,
J. Phys. Chem. C \textbf{113}, 9982 (2009).

\bibitem{Sepioni10} M. Sepioni, R. R. Nair, S. Rablen, J. Narajanan, F.
Tuna, R. Winpenny, A. K. Geim, and I. V. Grigorieva, Phys. Rev. Lett.
\textbf{105}, 207205 (2010).

\bibitem{Yazyev08} O. V. Yazyev and M. I. Katsnelson, Phys. Rev. Lett.
\textbf{100}, 047209 (2008).

\bibitem{Cervenka09} J. \v{C}ervenka, M. I. Katsnelson, and C. F. J. Flipse,
Nature Phys. \textbf{5}, 840 (2009).

\bibitem{Lin09} H.-H. Lin, T. Hikihara, H.-T. Jeng, B.-L. Huang, C.-Y. Mou,
and X. Hu, Phys. Rev. B \textbf{79}, 035405 (2009).

\bibitem{Yazyev07} O. V. Yazyev and L. Helm, Phys. Rev. B \textbf{75},
125408 (2007).

\bibitem{Kumazaki07} H. Kumazaki and D. S. Hirashima, J. Phys. Soc. Jpn.
\textbf{76}, 064713 (2007)

\bibitem{Pisani08} L. Pisani, B. Montanari, and N. Harrison, New J. Phys.
\textbf{10}, 033002 (2008).

\bibitem{Palacios08} J. J. Palacios, J. Fern\'{a}ndez-Rossier, and L. Brey,
Phys. Rev. B \textbf{77}, 195428 (R)(2008).

\bibitem{Yazyev10} O. V. Yazyev, Rep. Prog. Phys. \textbf{73}, 05650 (2010).

\bibitem{Hentschel07} M. Hentschel and F. Guinea, Phys. Rev. B \textbf{76},
115407 (2007).

\bibitem{Zhu10} Z.-G. Zhu, K.-H. Ding, and J. Berakdar, Eur. Phys. Lett.
\textbf{90}, 67001 (2010).

\bibitem{Fuhrer10} J.-H. Chen, W. G. Cullen, E. D. Williams, and M. S.
Fuhrer, arXiv\_1004.3373.

\bibitem{Uchoa10} B. Uchoa, T. G. Rappoport, and A. H. Castro Neto,
arXive:1006.2512v1.

\bibitem{Brey07} L. Brey, H. A. Fertig, and S. Das Sarma, Phys. Rev. Lett.
\textbf{99}, 116802 (2007).

\bibitem{Cheianov07} V. V. Cheianov, Eur. Phys. J. Special Topics \textbf{148%
}, 55 (2007).

\bibitem{Rappoport09} T. G. Rappoport, B. Uchoa, and A. H. Castro Neto,
Phys. Rev. B \textbf{80}, 245408 (2009).

\bibitem{Venezuela09} P. Venezuela, R. B. Muniz, A. T. Costa, D. M. Edwards,
S. R. Power, and M. S. Ferreira, Phys. Rev. B \textbf{80}, 241413(R) (2009).

\bibitem{Schaffer10} A. M. Black-Schaffer, Phys. Rev. B \textbf{81}, 205416
(2010).

\bibitem{Leib} E. H. Lieb, Phys. Rev. Lett. \textbf{62}, 1201 (1989).

\bibitem{Easquinazi10} P. Esquinazi, J.Barzola-Quiquia, D. Spemann, M.
Rothermel, H. Ohldag, N.Garc\'{\i}a, A. Setzer, and T. Butz, J. Magn. Magn.
Mat. \textbf{322}, 1156 (2010).

\bibitem{Lopez10} M. P. L\'{o}pez-Sancho , F. de Juan, and M. A. H.
Vozmediano, J. Magn. Magn. Mat. \textbf{322}, 1167 (2010).

\bibitem{Mermin66} N. D. Mermin and H. Wagner, Phys. Rev. Lett. \textbf{17},
1133 (1966).

\bibitem{Brataas06} D. Huertas-Hernando, F. Guennea, and A. Brataas, Phys.
Rev. B \textbf{74}, 155426 (2006).

\bibitem{Neto09} A. H. Castro Neto, F. Guinea, N. M. R. Peres, K. S.
Novoselov, and A. K. Geim, Rev. Mod. Phys. \textbf{81}, 109 (2009).

\bibitem{LP9} E. M. Lifshitz and L. P. Pitaevskii, \textit{Statistical
physics} (Pergamon, Oxford, 1980), Part~2.

\bibitem{Englman72} R. Englman, \textit{The Jahn-Teller Effect in Molecules
and Crystals} (Wiley, New York, 1972).

\bibitem{Bednarski00} H. Bednarski, J. Magn. Magn. Mat. \textbf{213}, 377
(2000).

\end{thebibliography}
\end{document}